\documentclass[twocolumn,prb]{revtex4}
\usepackage{graphicx}
\usepackage{amsmath}
\usepackage{amssymb}
\usepackage{epsf}
\usepackage{bm}

\renewcommand{\phi}{\varphi}
\renewcommand{\epsilon}{\varepsilon}
\renewcommand{\vec}[1]{{\bf #1}}

\usepackage{color}

\newcommand{\Avg}[1]{\langle  #1  \rangle}
\newcommand{\la}{\langle}
\newcommand{\ra}{\rangle}
\newcommand{\lb}{\left[}
\newcommand{\rb}{\right]}
\newcommand{\lp}{\left(}
\newcommand{\rp}{\right)}

\newcommand{\be}{\begin{equation}}
\newcommand{\ee}{\end{equation}}
\newcommand{\bea}{\begin{eqnarray}}
\newcommand{\eea}{\end{eqnarray}}

\newcommand{\p}{\partial}

\renewcommand{\phi}{\varphi}
\renewcommand{\epsilon}{\varepsilon}
\renewcommand{\vec}[1]{{\bf #1}}

\begin{document}

\title{Generating Entanglement and Squeezed States of Nuclear Spins in Quantum Dots}

\author{M. S. Rudner$^{1}$, L. M. K. Vandersypen$^{2}$, V. Vuleti\'{c}$^{3}$, L. S. Levitov$^{3}$}
\affiliation{
$^{1}$ Department of Physics, Harvard University, 17 Oxford St., 5 Cambridge, MA 02138\\
$^{2}$ Kavli Institute of NanoScience, TU Delft, PO Box 5046, 2600 GA, Delft, The Netherlands\\
$^{3}$ Department of Physics, Massachusetts Institute of Technology, 77 Massachusetts Ave, Cambridge, MA 02139
}

%\date{\today}

%\begin{abstract}

%\end{abstract}

\maketitle

{\bf  Entanglement generation and detection are two of the most sought-after goals in the field of quantum control.
Besides offering a means to probe some of the most peculiar and fundamental aspects of quantum mechanics, 
entanglement in many-body systems can be used as a tool to reduce fluctuations below the standard quantum limit.
For spins, or spin-like systems, such a reduction of fluctuations can be realized with so-called squeezed states. %\cite{Kitagawa}. 
Here we present a scheme for achieving coherent spin squeezing of nuclear spin states in semiconductor quantum dots.
This work represents a major shift from earlier studies in quantum dots, which have explored 
classical ``narrowing'' of the nuclear polarization distribution through feedback involving stochastic spin flips. %~\cite{Klauser,Giedke,Stepanenko,Rudner1,Rudner2,Danon}.
In contrast, we use the nuclear-polarization-dependence of the electron spin resonance (ESR) 
to provide a non-linearity which generates a non-trivial, area-preserving, ``twisting'' dynamics, analogous to that introduced %in M. Kitagawa and M. Ueda, {\it Phys. Rev. A} 47, 5138 (1993), 
by Kitagawa and Ueda, that squeezes and stretches the nuclear spin Wigner distribution without the need for nuclear spin flips.}
%\end{abstract}

%Entanglement is one of the most peculiar features of quantum mechanics, which serves as a lies at the heart of the fundamental differences between classical and quantum behavior.
%Within the quantum regime, 
% means to produce
%condition 

%Over the past several years, single-electron quantum dots have emerged as a powerful platform for studying quantum spin dynamics in a solid-state setting \cite{Hanson07}. 
Recently, squeezing of the collective spin state of many atoms\cite{Kitagawa} was achieved using atom-light or atom-atom interactions\cite{Appel2009,Schleier-Smith,Gross2010,Riedel2010}. 
The resulting reduced uncertainty can be used to enhance the performance of atomic clocks, and to achieve unprecedented precision of measurements in atomic ensembles. % performed on cold atom ensembles.
Similarly, the ability to generate squeezed states in a solid-state setting will open new avenues for quantum control of collective degrees of freedom at the nanoscale.  %many new possibilities for investigating quantum dynamics at the nanoscale.
In particular, due to the ubiquitous presence of randomly oriented and evolving  nuclear spins in nanoscale solid-state devices\cite{Petta05, Koppens06, Pfund07, Mikkelsen, Greilich, Latta, Press10}, future progress in spin-based information processing %technologies
 hinges on our ability to find ways of precisely controlling the dynamics of nuclear spins.

%%%%%%%%%%%%%%%%%%%%%%%%%%%%%%%%%%%%%%%%%%%%%%%%%%%%%%%%%%%%%%%%%%%
\begin{figure}[t]
\begin{center}\includegraphics[width=3.2in]{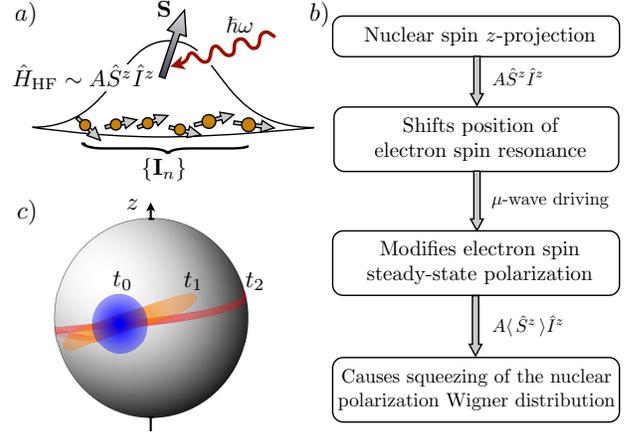}\end{center}
 \caption[]{Nuclear spin squeezing in a quantum dot. 
a) An electron in a quantum dot, with the electron spin $\vec{S}$ coupled to a large group of nuclear spins $\{\vec{I}_n\}$.
Electron spin resonance is excited by microwave radiation applied in the presence of an external magnetic field. % with frequency close to the Zeeman splitting. % in an applied magnetic field.
b) Flowchart describing the squeezing mechanism. % chain which leads to squeezing.
c) Schematic depiction of twisting dynamics on the Bloch sphere, shown in a rotating frame where the mean polarization is stationary. % with precession frequency $\eta_s$ a function of the nuclear spin $z$ projection, Eq.(\ref{eq:eta}). 
%The Wigner distribution is schematically shown before squeezing, $t_0$, and at two later times $t_{1,2}$. % as squeezing proceeds.
We focus on short to intermediate times $t_0 \lesssim t \lesssim t_1$, where
%Between times $t_0$ and $t_1$, 
the phase space (Wigner) distribution is squeezed within a small region of the Bloch sphere, with uncertainty decreasing as $1/t$ beyond the standard quantum limit, Eq.(\ref{dI_ideal}).
% acquiring reduced uncertainty.
% sub-Heisenberg uncertainty.
% and can be subsequently used for achieving sub-Heisenberg uncertainty.
% its reduced uncertainty.  
At longer times, indicated by $t_2$, the distribution extends around the Bloch sphere, at which point maximal achievable squeezing is reached.
%gets narrow still, but 
% becomes spiral-like as it extends around the Bloch sphere. %, and is therefore less useful as a reduced-uncertainty resource.
For moderate initial nuclear polarizations $p=2I/N$, significant squeezing can be achieved on the intermediate time scale $t\lesssim t_1$ (see text).
%distribution extends significantly around the Bloch sphere
%{\bf between $t_1$ and $t_2$ distribution gets narrower, but spirally and useless. however, there exists enough range to get useful squeezing}
%Dynamics shown for three different ratios of the driving amplitude $\Delta$ and maximal Overhauser field $\omega_{\rm overhauser}=AI_{\rm max}$.
%{\bf LV: Need to explain in captions that rotating frame is used where center of distribution is stationary.}
 }
\label{fig1}
\end{figure}
%%%%%%%%%%%%%%%%%%%%%%%%%%%%%%%%%%%%%%%%%%%%%%%%%%%%%%%%%%%%%%%%%%%

Recently developed experimental techniques provide a powerful toolkit for probing and controlling nanoscale groups of nuclear spins. %in nanoscale systems.
The electron Zeeman splitting offers a sensitive way to detect the collective nuclear spin polarization through the Overhauser shift of the electronic spin states. 
This splitting can be measured accurately using spectroscopic\cite{Koppens06, Latta} or time-domain measurement techniques\cite{ Mikkelsen, Greilich, Press10,  Bluhm}. %\cite{Petta05, Koppens06}. 
Furthermore, the hyperfine coupling provides a means to %also enables us to 
control the nuclei through their interactions with electron spins. 
Using such means, the Tarucha group has reported nuclear spin polarizations of up to 40$\%$~\cite{Baugh}. %, and polarizations of the order of 5$\%$ are realized routinely \cite{Koppens05,Marcuslabpaper}. 
Using optical pumping,
even larger polarizations of up to $60\%$ have been demonstrated \cite{Gammon}. %In addition, %the randomness 
It was proposed that fluctuations of the nuclear polarization can be ``narrowed'' through various schemes exploiting electron-nuclear feedback through controlling stochastic spin-flips of the nuclei~\cite{Klauser,Giedke,Stepanenko,Rudner1,Rudner2,Danon}.
%Fluctuations in one component of the nuclear spin polarization have also been suppressed by 
Recently, such methods have been successfully used to achieve narrowing by a factor of 5-10 in several systems~%, by exploiting electron-nuclear feedback
\cite{Greilich, Latta,  Bluhm,Vink,Steel, Ladd10}. 
These techniques, together with the possibility of %Using externally applied RF fields, 
driving coherent nuclear spin rotations using externally applied RF fields\cite{Machida1,Machida2,Hirayama,Tartakovskii,Takahashi10}, 
can be used to explore a wide range of quantum phenomena with nuclear spins.
%provide us with a powerful toolkit for controlling nuclear spins in nanoscale devices. %quantum dots.  %of the polarization of a local ensemble of nuclear spins in a nanostructure have been realized as well\cite{Machida,Hirayama,Tartakovskii}.

Here we describe a spin squeezing mechanism that can be employed in a variety of systems such as spins controlled by microwave radiation in quantum dots defined by electrostatic gates in two-dimensional electron gases\cite{Koppens06}, mesas\cite{vanderWiel06}, or nanowires\cite{Nadj-Perge10}, or optically-controlled spins in self-assembled quantum dots or nanoparticles\cite{Mikkelsen, Greilich, Latta, Press10, Gammon}.
%While the experimental details may differ between different setups, the situation appears promising for all possibilities.
For concreteness, below we provide realistic estimates where appropriate using parameters relevant for gate-defined quantum dots in GaAs, but we emphasize that the situation appears similarly promising for these other types of quantum dot systems.

\section{Spin Squeezing in Quantum Dots}
In a system comprised of many spins $\{\hat{\vec{I}}_n\}$, such as a quantum dot or an atomic ensemble, 
the collective total spin $\hat{\vec{I}} = \sum_n \hat{\vec{I}}_n$ is a quantum mechanical angular-momentum variable.
Because different vector components of $\hat{\vec{I}}$ do not commute, they are subject to the Heisenberg uncertainty relations
\be\label{HUP}
\Delta I^y \Delta I^z \ge
\frac{\hbar}{2}|\Avg{\hat{I}^x}|,
\ee
and its cyclic permutations,
where $\Delta I^\alpha = \Avg{(\delta \hat{I}^\alpha)^2}^{1/2}$, with $\delta \hat{I}^\alpha = \hat{I}^\alpha - \Avg{\hat{I}^\alpha}$.
The goal of squeezing is to reduce fluctuations in one spin component below the ``quantum limit,'' $\Delta I^\alpha = \sqrt{\frac12\hbar|\Avg{\hat{I}^x}|}$, set by Eq.(\ref{HUP}) in the situation where %the microscopic spins are uncorrelated and 
the fluctuations are distributed equally between both orthogonal components.
Quantitatively, a state is therefore said to be quantum mechanically squeezed along an axis $z$ when the squeezing parameter\cite{Kitagawa, Wineland} 
\bea
\label{SqueezingCond}\xi = \frac{\Delta I^z}{\sqrt{\frac12 \hbar |\Avg{\hat{I}^x}|}} %\sqrt{\frac{2\, (\Delta I^z)^2}{\hbar |\Avg{\hat{I}^x}|}} %\frac{\sqrt{2I}\,\Delta I^z}{|\Avg{\hat{I}^x}|}
\eea
is less than one.
%{\bf I tried to formulate the condition without explicit reference to coherent states with fixed length. Are we okay with this usage?}
Because the Heisenberg uncertainty principle must remain satisfied, a reduction of fluctuations $\Delta I^z$ signified by $\xi < 1$ must be accompanied by an excess of fluctuations $\Delta I^y$ above the quantum limit.
% in In the quantum limit, where the inequality in Eq.(\ref{HUP}) is nearly saturated, fluctuations in one spin component can be reduced only at the expense of increased fluctuations in another component.

In contrast to the situation in atomic ensembles where classical fluctuations of the collective total spin magnitude $\hat{\mathbf{I}}^2$ can be negligible, however, additional fluctuations in nuclear spin states in quantum dots are unavoidable due to the abundance of nearly degenerate states with energy splittings much smaller than the temperature. 
Therefore, for typical states in quantum dots, the uncertainty relation (\ref{HUP}) is far from being saturated.
 In a typical situation where the equilibrium state of $N \approx 10^6$ microscopic nuclear spin-1/2 moments is completely random, the classical uncertainties $\Delta I^y = \Delta I^z = \sqrt{N}\hbar/2$ %${\Avg{(\delta \hat{I}^y)^2}}$ and ${\Avg{(\delta \hat{I}^z)^2}}$ 
are identical in magnitude to the quantum uncertainty % those present due to ``projection noise'' 
in the maximally polarized (product) state with $\Avg{\hat{I}^x} = \hbar N/2$, see Eq.(\ref{HUP}).%, where $N$ is the number of microscopic spin-1/2 moments comprising the total spin.
\footnote{For simplicity we consider microscopic spins of size $s = 1/2$, but the generalization to other spins is straightforward.}
Below we consider an initial state prepared by polarizing nuclear spins to a fraction $p$ of the maximal polarization, and then rotating this polarization into the equatorial plane of the Bloch sphere such that the mean spin points along $x$, $\Avg{\hat{I}^x}_0 = pN\hbar/2$.
%assume that {\bf consider initial state prepared by polarizing and rotating to xy plane}, while nuclear spins cannot be fully polarized, through some mechanism of dynamical nuclear polarization (DNP) they are initially polarized to a fraction $p$ of the maximal polarization, $\Avg{\hat{I}^x} = pN\hbar/2$.
%\mpar{flucts.~in length of $I$?}
%Using Eq.(\ref{HUP}), 
%Again using Eq.(\ref{HUP}), the isotropic quantum mechanical minimum fluctuations in such a state are given by $\Delta I^y = \Delta I^z \approx \hbar\sqrt{fN/2}$.
Because the classical fluctuations in the orthogonal components of the initial state $\Delta I_0^{y,z}$, which are not affected by dynamical nuclear polarization (DNP), are only a factor of $\frac1{\sqrt{p}}$ larger than the quantum mechanical  minimum fluctuations, $\Delta I^{y,z} \approx \sqrt{pN}\hbar/2$, %$\Delta I^y = \Delta I^z \approx \sqrt{fN\hbar/2}$ set by Eq.(\ref{HUP}). %$\sqrt{\Avg{(\delta \hat{I}^y)^2}} = \sqrt{\Avg{(\delta \hat{I}^z)^2}} \approx \sqrt{\hbar fN/2}$ 
%of such a polarized state, % are reduced only by a factor of $\sqrt{f}$ compared to the uncertainty in the random initial state, 
%Hence 
only a modest amount of additional overhead is required in order to achieve squeezing down to the quantum limit.
%{\bf Do we care about bringing one component below the Heisenberg limit, or saturating the uncertainty relation?}
% Quantitatively, when the squeezing parameter\cite{Kitagawa, Wineland}
% \bea
% \label{SqueezingCond}\xi = \frac{\Delta I^z}{\sqrt{\frac12 \hbar |\Avg{\hat{I}^x}|}} %\sqrt{\frac{2\, (\Delta I^z)^2}{\hbar |\Avg{\hat{I}^x}|}} %\frac{\sqrt{2I}\,\Delta I^z}{|\Avg{\hat{I}^x}|}
% \eea
% is less than one, then the fluctuations in one spin component (here taken to be $z$) have been reduced below the standard quantum limit, and the state is said to be quantum mechanically squeezed.
% {\bf I tried to formulate the condition without explicit reference to coherent states with fixed length. Are we okay with this usage?}

We note that, for any mechanism of DNP, the extent of polarization $p$ itself will contain fluctuations.
%Care must be taken to check that such fluctuations will not pose a problem for squeezing. 
Below we will check, and confirm, that such fluctuations do not present any significant additional challenges for squeezing.

In contrast to recent experiments which achieved spin squeezing through atom-light or atom-atom interactions \cite{Appel2009,Schleier-Smith,Gross2010,Riedel2010}, in the mechanism that we outline below, the nuclear spins are driven through their interaction with a single electron spin.
As we discuss, the effective squeezing Hamiltonian for nuclei is produced after motional-averaging of the quickly fluctuating, strongly driven electron spin.
%time-averaged {\it effective field} of a single electron spin. %produced by the motion-averaged directly through their interaction with a single spin-1/2 electron.
While the end results (squeezed collective spin states) of these approaches are similar, the physical mechanisms are quite different.
%Whereas phase diffusion due to photon shot noise in the atomic case can be reduced by working in the regime of large cavity photon number, here phase diffusion can be suppressed by making use of ``motional-averaging'' in the strong driving regime where the electron spin fluctuates rapidly between the up and down spin states. 
%As we will discuss below, here fluctuations about the average

\section{The Model: Squeezing Dynamics}
The underlying physics of the feedback mechanism can be understood most easily in the regime of fast dephasing of the electron spin, which is also a relevant regime for current experimental efforts. 
As depicted in Fig.~\ref{fig1}a, we consider a single electron in a quantum dot, %subject to an in-plane external magnetic field, 
in contact with a large group of nuclear spins, $\{\hat{\vec{I}}_n\}$. %, in a quantum dot.
The electron and nuclear spins are coupled by the hyperfine interaction $H_{\rm HF} = \sum_n A_n \hat{\vec{S}}\cdot\hat{\vec{I}}_n$, where $\hat{\vec{S}}$ is the electron spin, and each coupling constant $A_n$ is proportional to the local electron density at the position of nucleus $n$.
%We shall see that under these conditions, the electron-nuclear hyperfine coupling can lead to an effective Hamiltonian for the nuclear spins $\hat {\vec I}=\sum_{i=1...N}\hat {\vec s}_{i}$. 
The electron spin %\mpar{mention $B_{\rm ext}$ here?} %, which is subjected to an in-plane external magnetic field, 
is driven by an applied RF field with frequency close to the electron spin resonance in the presence of an externally applied magnetic field. 
Because the electron spin evolves rapidly on the timescale of nuclear spin dynamics, the nuclear spins are subjected to an effective hyperfine field (the ``Knight field'') produced by the time-averaged electron spin polarization.
Feedback results from the dependence of the average electron spin polarization on the detuning from the ESR condition, which in turn depends on the nuclear polarization, see Fig.~\ref{fig1}b.

To describe this regime, we model the system with the microscopic Hamiltonian (here and below we take $\hbar = 1$)
\be
\label{Hmicro} H = \omega_Z \hat{S}^z + \omega_0 \hat{I}^z + A \hat{I}^z\hat{S}^z + \frac{A}{2}(\hat{I}^+\hat{S}^- + \hat{I}^-\hat{S}^+) + H_{\rm el},
\ee 
where $\omega_Z$ is the electron Zeeman energy in the magnetic field, $\omega_0$ is the nuclear Larmor frequency, and $H_{\rm el}$ describes the driving of the electron spin and its coupling to an environment, which leads to fast dephasing and relaxation.
For simplicity, here we consider a single species of nuclear spin, and take all hyperfine coupling constants to be equal, $A_n = A$.
The latter condition amounts to the assumption that electron density is approximately constant inside the dot, and zero outside. 
In this case, the electron spin couples directly to the {\it total} nuclear spin $\hat{\vec{I}} = \sum_n \hat{\vec{I}}_n$, with the square of the total nuclear spin, $\hat{I}^2$, conserved by the dynamics.
The effects of non-uniform couplings will be discussed at the end.
%For simplicity, we take the hyperfine coupling $A$ to be the same for all nuclear spins in the dot.
%As a result, the electron spin couples directly to the {\it total} nuclear spin $\vec{I} = \sum_n \vec{I}_n$, where $n$ labels individual nuclear spins, with the square of the total nuclear spin, $I^2$, conserved by the dynamics.
%In this case, the nuclear spins act together collectively as one ``giant spin'' of conserved length, which couples to the electron.
%We will discuss modifications to this picture due to non-uniform couplings later. %{\bf Some words about giant spin.  Will discuss effects of non-uniform $A$ later.} 
Due to the large mismatch between the electron and nuclear Zeeman energies, $\omega_Z/\omega_0 \gg 1$, below we ignore the ``flip-flop'' terms proportional to $\hat{I}^+ \hat{S}^-$  and $\hat{I}^-\hat{S}^+$ in Eq.(\ref{Hmicro}).

%%%%%%%%%%%%%%%%%%%%%%%%%%%%%%%%%%%%%%%%%%%%%%%%%%%%%%%%%%%%%%%%%%%
\begin{figure}[t]
\begin{center}\includegraphics[width=3.2in]{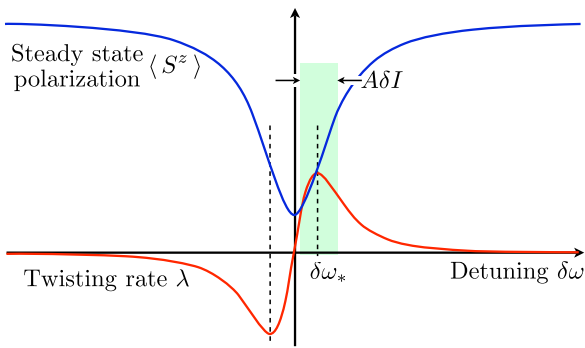}\end{center}
 \caption[]{Time-averaged electron spin polarization $S^z$, Eq.(\ref{eq:S_z}), and squeezing strength $\lambda$, Eq.(\ref{eq:H_Kitagawa}), versus RF detuning $\delta\omega$ from the ESR frequency.
The average electron spin polarization depends on ${\hat I}^z$ through the dependence of the detuning on the Overhauser shift, as indicated by the shaded region.
Squeezing is most efficient for the detuning $\delta\omega_* = \tilde\gamma/\sqrt{3}$ where $S^z$ is most sensitive to ${\hat I}^z$, see text.
 }
\label{Lorentzian}
\end{figure}
%%%%%%%%%%%%%%%%%%%%%%%%%%%%%%%%%%%%%%%%%%%%%%%%%%%%%%%%%%%%%%%%%%%
We begin by writing the Heisenberg equation of motion for the total nuclear spin operator $\hat{\vec{I}}$, $d\hat {\vec I}/dt=i[\hat {\vec I},H]$:
\be\label{eq:bloch_eqs}
\frac{d\hat{\vec I}}{dt}=\vec b \times\hat{\vec I}
,\quad
\vec b= \lp \omega_0 + A \hat{S}^z \rp  \vec z.
\ee
When electron dephasing is fast on the characteristic time scale of nuclear spin dynamics, ``motional-averaging'' allows the electron polarization $\hat{S}^z$ in %to be eliminated from
Eq.(\ref{eq:bloch_eqs}) to be replaced by an operator-valued semiclassical mean polarization $S^z(\hat{I}^z)$ which depends on the nuclear polarization $\hat{I}^z$ through the Overhauser shift of the ESR frequency, c.f. Ref.\cite{Schelier-SmithPRA}.
%we eliminate the dependence of Eq.(\ref{eq:bloch_eqs}) on the electron polarization $\hat{S}_z$
%the electron spin polarization can be treated as a classical variable given by its ``motion-averaged'' expectation value $S_z=\la \hat S_z\ra$. 
%This semi-classical approach is supported by a more rigorous calculation involving the evolution of the full system's density matrix, to be presented elsewhere\cite{Squeezing2}.
%Note that the field $S_z$ which acts back on the nuclear spin depends on the nuclear polarization $I_z$ through the Overhauser shift of the ESR frequency. 
% Here the nuclear spin operators evolve according to the Heisenberg equations of motion $d\hat {\vec I}/dt=i[\hat {\vec I},H]$, which can be written as a Bloch equation for the expectation values $\vec I=\la \hat {\vec I}\ra$: 
% %
% \be\label{eq:bloch_eqs}
% \frac{d\vec I}{dt}=\vec b \times\vec I
% ,\quad
% \vec b= \lp \omega_0 + A \la S_z\ra \rp  \vec z,
% \ee
% %
% where $\la S_z\ra$ depends on $I_z$ through the dynamical Overhauser shift of the ESR frequency. 
In this %fast dephasing 
regime, %the electron spin dynamics and 
%steady-state mean-field polarization 
${S}^z(\hat{I}^z)$ %, which depend on the nuclear polarization $I_z$ through the Overhauser shift of the ESR frequency, 
%are described simply
%Because we focus here on the regime of fast dephasing, electron spin dynamics can be simply described 
%in terms of 
is simply determined by rate equations involving the occupation probabilities $n_+$, $n_-$ of the up and down electron spin states:
\bea\label{eq:rate_eqs}
&& \dot n_-=W(n_+-n_-)-\Gamma_1 n_-
\\\nonumber 
&& \dot n_+=W(n_--n_+)+\Gamma_1 n_-.
\eea
Here $W=\frac12\Omega^2\gamma/[(\delta\omega-A\hat{I}^z)^2+\gamma^2]$ is the ($\hat{I}^z$-dependent) %electron spin resonance (ESR) 
ESR transition rate with  detuning $\delta\omega$ between the driving frequency and $\omega_Z$, driving strength $\Omega$, and electron spin dephasing rate $\gamma \equiv 1/T_2$, while $\Gamma_1$ is the electron spin relaxation rate.
%Note that $\gamma \ge \frac12\Gamma_1$.
%Electron spin relaxation te $\Gamma_1$. % is the electron spin (up-down) relaxation rate.
%Assuming that the ESR line is broadened by intrinsic dephasing, the transition rate $W$ has a Lorentzian line shape $W=\frac12\Omega^2\gamma/[(\delta\omega-AI_z)^2+\gamma^2]$ with characteristic width $\gamma$.
%\mpar{changed $(\gamma/2)^2$} %which describes an ESR resonance broadened by intrinsic dephasing, yields Eq.(\ref{eq:S_z}).
%Here $\Omega$ is the driving field strength, $\gamma = 1/T_2$ is the electron spin dephasing rate, and $\delta \omega$ is the detuning between $\omega_Z$ and the driving frequency.
%Combined with Eq.(\ref{eq:rate_eqs}), this gives
Using this form for $W$, the steady-state solution of Eq.(\ref{eq:rate_eqs}) gives
\be\label{eq:S_z}
S^z =\frac12\,\frac{(\delta\omega-A\hat{I}^z)^2+\gamma^2}{(\delta\omega-A\hat{I}^z)^2+\tilde\gamma^2}
,\quad
\tilde\gamma^2=\gamma^2+\frac{\gamma}{\Gamma_1}\Omega^2,
\ee
where $\tilde{\gamma}$ includes the effect of power broadening. % {\bf [Ref]}.
%where $\Omega$ is the driving field strength, $\gamma$ is the electron dephasing rate, and $\Gamma_1$ is the up-down relaxation rate. 

% This form of $\la S_z\ra$ arises naturally from the rate process describing the system in terms of
% the occupation probabilities of the up and down electron spin states,
% %
% \bea\label{eq:rate_eqs}
% && \dot n_-=W(n_+-n_-)-\Gamma_1 n_-
% ,
% \\\nonumber 
% && \dot n_+=W(n_--n_+)+\Gamma_1 n_-
% .
% \eea
% %
% For a steady state we have 
% %
% \be
% \la S_z\ra=\la\frac12(n_+-n_-)\ra=\frac12\frac{\Gamma_1}{2W+\Gamma_1}.
% \ee
%

\section{Results: Squeezing Rate and Efficiency}
The connection with nuclear spin squeezing is elucidated by linearizing Eq.(\ref{eq:S_z}) in $A\hat{I}^z$ around the optimal detuning $\delta\omega_* = \tilde\gamma/\sqrt{3}$ where $S^z$ is most sensitive to nuclear-polarization-dependent frequency shifts, see Fig.\ref{Lorentzian}.
Here the nuclear Bloch dynamics described by Eq.(\ref{eq:bloch_eqs}) can be mapped onto one of the canonical squeezing Hamiltonians proposed by Kitagawa and Ueda\cite{Kitagawa}: % when the Overhauser shift $AI_z$ is small compared to the `combined' broadening $\tilde\gamma$, Eq.(\ref{eq:S_z}).
%In this regime, our problem is reduced to that considered in Ref.\onlinecite{Kitagawa} by linearizing $\la S_z\ra$ in small $AI_z$ on the shoulder of the Lorentzian profile, see Eq.(\ref{eq:S_z}) and Fig.\ref{Lorentzian}.
absorbing a constant term into the net nuclear Larmor frequency $\omega_0$, we obtain 
\be\label{eq:H_Kitagawa}
H\approx \omega_0 \hat{I}^z + \frac12 \lambda (\hat{I}^z)^2
,\quad
\lambda=A \left.\frac{\p S^z}{\p \hat{I}^z}\right\vert_{\begin{subarray}{l}\hat{I}^z = 0\\\delta\omega=\delta\omega_*\end{subarray}},
\ee
with $\hat I^2=I(I+1)$, $I \le N/2$, conserved by the dynamics. 

%%%%%%%%%%%%%%%%%%%%%%%%%%%%%%%%%%%%%%%%%%%%%%%%%%%%%%%%%%%%%%%%%%%
\begin{figure}[t]
\begin{center}\includegraphics[width=3.2in]{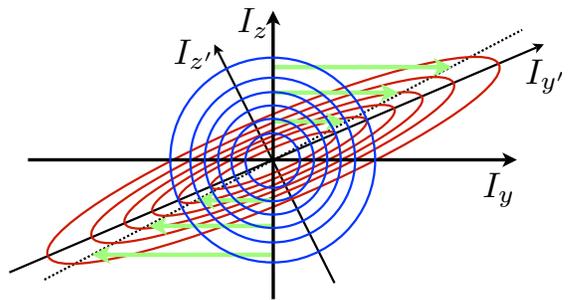}\end{center}
 \caption[]{Contour plot representation of the Wigner distribution of a large collective spin on a locally-flat patch of the  Bloch sphere, in the rotating frame where $\omega_0 = 0$. The mean spin points along $x$.
This picture applies for times  $t_0 \le t \le t_1$, see Fig.\ref{fig1}c.
Before squeezing, the Wigner distribution is isotropic (blue).
After squeezing, the Wigner distribution, Eq.(\ref{eq:evolved_gaussian}), is squeezed along an axis $z'$, and stretched along an orthogonal axis $y'$ (red).
Within the mean-field approximation, phase space area is preserved. 
%{\bf Short time before distribution spread over Bloch sphere ($t_1$ and not $t_2$ in fig 1c)}
%Time-averaged electron spin polarization $\Avg{S^z}$ and squeezing strength $\lambda$ versus RF detuning $\delta\omega$ from the ESR frequency.
%The average electron spin polarization depends on $I^z$ through the dependence of the detuning on the Overhauser shift, as indicated by the shaded region.
%Squeezing is most efficient for the detuning $\delta\omega_* = \tilde\gamma/2\sqrt{3}$ where $\Avg{S^z}$ is most sensitive to $I^z$, see text.
 }
\label{Planar}
\end{figure}
%%%%%%%%%%%%%%%%%%%%%%%%%%%%%%%%%%%%%%%%%%%%%%%%%%%%%%%%%%%%%%%%%%%
The squeezing induced by Hamiltonian (\ref{eq:H_Kitagawa}) is most simply illustrated by the evolution of a coherent nuclear spin state of length $I$ initially oriented along $x$. 
For large $I$, semiclassical analysis shows that the quadratic term in Eq.(\ref{eq:H_Kitagawa}) %can be understood semiclassically as causing
induces precession of the total spin vector about the $z$ axis with a {\it polarization-dependent} Larmor frequency $\eta=\partial H/\partial I^z=\omega_0 +\lambda I^z$.
%dynamics generated by Eq.(\ref{eq:H_Kitagawa}) can be understood semiclassically as precession of the total spin vector about the $z$ axis with the Larmor frequency $\eta=\partial H/\partial I_z=\omega_0 +\lambda I_z$. Because the Larmor frequency is a function of the spin $z$ component, the dynamics on a Bloch sphere is of a twisting character, with linearly varying precession rate:
%
%\be\label{eq:twisting}
%I_x(t)+iI_y(t)\propto e^{i(\omega_0 +\lambda s) t}
%,\quad
%I_z=s
%.
%\ee
%
%This defines an area-preserving dynamics under which an area element on the Bloch sphere is stretched in one direction and squeezed in another direction. 
%{\bf Discuss twisting dynamics in small patch of Bloch sphere, mapped to the plane.  Show figure.}
%As illustrated in Fig.\ref{fig1}a, 
%As illustrated in Fig.\ref{fig1}c, t
The resulting evolution %on the Bloch sphere 
has a ``twisting'' character,
%Equation (\ref{eq:twisting}) defines an area-preserving ``twisting'' dynamics under which
in which an area element on the Bloch sphere is stretched in one direction and squeezed in another, see Fig.\ref{fig1}c. 
%Twisting dynamics of the form (\ref{eq:twisting}) can be realized in cold atoms coupled to light, leading to collective squeezed states \cite{Schleier-Smith}. 
%As discussed below Eq.(\ref{HUP}), the uncertainty relations prevent %presence of quantum (and classical) fluctuations prevents all three components of the nuclear polarization from being simultaneously well-defined.
Squeezing becomes significant at times $t\gtrsim t_S=(|\lambda| I)^{-1}$ when the relative precession between the upper and lower edges of the corresponding ``uncertainty region'' becomes comparable to the width of the region. % (see Supplementary Information).
Using Eq.(\ref{eq:S_z}), and assuming that the spread of Overhauser shifts $A\Delta I^z$ is small compared to the width of the resonance $\tilde\gamma$ to ensure the validity of the linearization, we find % that squeezing is effective for times
\be\label{t_min}
t_S \approx \frac{16\Gamma_1\tilde\gamma^3}{3\sqrt{3}IA^2\gamma\Omega^2}.
%\frac{\Gamma_1}{\gamma}\frac{[\delta\omega^2+(\tilde\gamma/2)^2]^2}{A^2 \Omega^2 I \delta\omega}\approx\frac{2\Gamma_1\tilde\gamma^3}{3\sqrt{3}IA^2\gamma\Omega^2}
\ee

For an order-of-magnitude estimate of the squeezing time, we set $\Omega = \Gamma_1 = \frac15\gamma$.
This choice selects the regime of moderately strong electron spin dephasing where the resonance is broader than the minimum value $\gamma = \frac12\Gamma_1$.
In this practically relevant regime, the rate equations (\ref{eq:rate_eqs}) and the motional-averaging approximation can be safely applied.
%$\Omega \approx \Gamma_1 = \frac14 \gamma$ and find  $t_S \approx 4\gamma/(IA^2)$. %4.35
%all relaxation/dephasing rates equal, $\gamma\approx \Gamma_1\approx\Omega$, giving $t_{\rm *}\approx 4\gamma/(IA^2)$. 
Taking the `intrinsic' width of the resonance to be twice larger than the %comparable to the 
typical Overhauser field fluctuations,  $\gamma\approx A\sqrt{N}$, we obtain
%{\bf Adjust ratios of parameters to be consistent with $a \ll b$, etc.  Lieven: discuss why $T_1$ so short (10 ns), when naturally expect seconds.}
%
\be
t_{S, {\rm min}}\approx 20\frac{\sqrt{N}}{IA}.
\ee
Using a typical value of the hyperfine coupling for GaAs, $A\approx 0.1 \,{\rm \mu s}^{-1}$, we obtain $t_{S,{\rm min}} \approx 200\, \mu s(\sqrt{N}/I)$.
%For partially polarized initial states, with $I/\sqrt{N} \gg 1$, the squeezing rate can be enhanced considerably. 
% and for an estimate took the Bloch vector length corresponding to unpolarized spins, $I=\sqrt{N}$. 
The estimate for $t_{S,{\rm min}}$ can be improved slightly by optimizing expression (\ref{t_min}) with respect to driving power $\Omega$. 
The fast relaxation rate $\Gamma_1 \sim A\sqrt N$ can be achieved by working in a regime of efficient electron spin exchange with the reservoirs in the leads.
We see that the squeezing time is inversely proportional to the initial length of the nuclear spin vector, i.e. the degree of nuclear polarization before squeezing.
%More importantly, we see that  however, a large improvement can be achieved by working with spins which are polarized before squeezing, $I\gg\sqrt{N}$. %, yields a shorter squeezing time. 
%{\bf Lieven: discuss realistic parameters for experiments?}

\section{Discussion: Effects of Fluctuations}

To investigate the effects of classical fluctuations in %all three components of the %the length and in the transverse components of the 
the initial state, as well as the effect of time-dependent electron spin fluctuations around the steady state $S^z$, we %obtain a quantitative picture of squeezing by 
now analyze the evolution of the nuclear spin Wigner distribution. 
%This approach will also allow us to analyze the effect of time-dependent electron spin fluctuations around the steady state $S^z$.
%First we consider the effect of classical fluctuations in the transverse components of the initial polarization.
For a large initial polarization $p$, where $I = pN/2$, and for short to intermediate times $t_0 \lesssim t \lesssim t_1$ (see Fig.\ref{fig1}c), the ``uncertainty region'' associated with the nuclear state is small on the scale of the total spin and we can consider evolution in a locally flat patch of the Bloch sphere. % (see Fig.\ref{Planar}).
Here the operators $\hat{I}^y$ and $\hat{I}^z$ approximately obey canonical commutation relations, and the initial nuclear spin state (polarized along $x$) is described by an isotropic 2D Gaussian Wigner distribution with width set by the initial transverse fluctuations, $\Delta I \equiv \Delta I^{y,z}_0$.
%,  $f(I_y,I_z)= \mathcal{A} e^{-(I_z^2+I_y^2)/2\delta I^2}$, with $\delta I = \Delta I_0^{y,z}$.
%{\bf talking about between $t_0$ and $t_1$ in Fig.1c}
Under the polarization-dependent precession generated by Eq.(\ref{eq:H_Kitagawa}), the Gaussian Wigner distribution evolves to
\be\label{eq:evolved_gaussian}
f_t(I^y,I^z)=\mathcal{A} \exp\lp -\frac{(I^z)^2+(I^y+I \lambda t I^z)^2}{2\Delta I^2}\rp,
\ee
where without loss of generality we set $\omega_0=0$. 
% Equal probability contours of t
The initial (isotropic) and evolved (squeezed) distributions are shown in Fig.\ref{Planar}.

The quadratic form in the exponential in Eq.(\ref{eq:evolved_gaussian}) is diagonalized in a suitably chosen orthonormal basis $y'$, $z'$ (see Supplementary Material). 
%, giving
%%
%\be\label{eq:wigner_twisting}
%f_t(I_z,I_y)=A \exp\lp -\frac{\cot^2\phi_t I_{z'}^2+\tan^2\phi_t I_{y'}^2}{2\delta I^2}\rp
%,
%%\label{eq:stretching_factor}
%% w =\sqrt{1+\frac{(\lambda I t)^2}{4}}+\frac{\lambda I t}{2}
%% w_t =\cot\phi_t 
%% ,\quad 
%\ee
%
%where $\cot 2\phi_t=\lambda I t/2$. 
As shown in Fig.\ref{Planar} and in Eq.(S5) of the Supporting Information, %(\ref{eq:wigner_twisting}), 
stretching in one direction ($y'$) %$\cot \phi_t$ times 
is accompanied by squeezing in the perpendicular direction ($z'$), such that the %which exactly preserves the 
%by the same amount, preserving the 
phase space volume of the Wigner distribution is exactly preserved if fluctuations of the electron spin are ignored. %\mpar{remove emphasis on area preservation?} 
For times $t \gtrsim t_S$, see Eq.(\ref{t_min}), the uncertainty $\widetilde{\Delta I}$ % \propto 1/(\lambda I t)$ 
of the squeezed component decreases as
\be\label{dI_ideal}
\widetilde{\Delta I}(t) \approx %\delta I\frac{2\Gamma_1\tilde\gamma^3}{3\sqrt{3}IA^2\gamma\Omega^2t} = 
\Delta I \frac{t_S}{t}.
\ee
%{\bf Discuss no limit to maximum squeezing ratio in the plane.  On the sphere, curvature limits maximum squeezing. Over long times get spiral.  Show figure.  Increasing $I$ with $\delta I \propto \sqrt{I}$ allows higher squeezing ratio on sphere. Estimate limit.}
%\mpar{put somewhere} For short times, Eq.(\ref{eq:evolved_gaussian}) for the Wigner distribution in a locally flat patch of the Bloch sphere gives a good description of the squeezing dynamics.
%However, as illustrated in Fig.\ref{fig1}b), at longer times the curvature of the Bloch sphere becomes important.
%Once the Wigner distribution begins to wrap around the Bloch sphere, the squeezing no longer translates into a simple reduction of fluctuations for one spin component. 
% stretched over distances comparable to the 
%Thus the curvature of the Bloch sphere imposes a limit on the maximum achievable squeezing, which scales as the ratio of the radius of the initial uncertainty cloud to the radius of the Bloch sphere, $\delta I/I$.
%For a state with typical uncertainty $\delta I \propto \sqrt{N}$, the maximum squeezing ratio goes as $I/\sqrt{N}$.
%This result can be applied until 
Squeezing proceeds until long times when the phase space distribution begins to extend around the Bloch sphere, see Fig.\ref{fig1}c.
The curvature of the Bloch sphere imposes a limit on the maximum achievable squeezing.\cite{Kitagawa}
%, which scales as the ratio of the radius of the initial uncertainty cloud to the radius of the Bloch sphere, $\Delta I/I$.\cite{Kitagawa}
%For a state with typical uncertainty $\delta I \propto \sqrt{N}$, the maximum squeezing ratio goes as $I/\sqrt{N}$.

To derive the squeezing time $t_S$ in Eq.(\ref{t_min}), a coherent nuclear spin state with $\Delta I_{\rm CSS} = \sqrt{I/2}$ was used.
As discussed above, however, when classical uncertainty in the nuclear spin state is included, the initial width of the Wigner distribution %initial transverse fluctuations are 
is given by $\Delta I = \sqrt{N}/2$.
Given that the width $\widetilde{\Delta I}$ of the squeezed component decays as $1/t$, see Eq.(\ref{dI_ideal}), the effect of the classical transverse fluctuations %in the initial state 
is simply to increase the time required to reach a desired level of fluctuations by an order-one factor $\sqrt{N/2I}=\frac1{\sqrt{p}}$. % $\sqrt{3^{1/2}N/2[I(I+1)]^{1/2}}$.

Besides fluctuations in the transverse components of the initial polarization, the DNP process used to prepare the initial nuclear spin state will also leave behind uncertainty in the length $I$ of the net spin (typically with a scale much smaller than $I$ itself).
Because Eq.(\ref{eq:evolved_gaussian}) describes {\it angular precession} with a rate which depends only on the $z$-component of the total spin, however, sections of the phase space distribution with constant $I^z$ but varying Bloch sphere radii $I$ rigidly precess without growing.
Therefore fluctuations in the initial polarization $I$ do not pose a significant threat to squeezing.

In addition to uncertainty in the initial nuclear spin state, we must also consider the effect of time-dependent fluctuations of the electron spin about its mean-field value $S^z$, Eq.(\ref{eq:S_z}).
The mean-field approximation to Eq.(\ref{eq:bloch_eqs}) applies in the motion-averaged limit when the %dynamics of the electron spin is fast
electron spin evolves quickly on the time scale of the nuclear spin dynamics, and hence the contribution of time-dependent electron spin fluctuations is small.
The residual effect of such fluctuations is to add a diffusive component to the nuclear-polarization-dependent precession induced by the time-averaged electron spin.
As shown in the Supplementary Material, the diffusivity $\kappa$ associated with this phase diffusion approximately goes as $\kappa \sim 1/\Gamma$, where $\Gamma \sim W,\Gamma_1$ is the characteristic rate of electron spin dynamics.
Thus phase diffusion is indeed suppressed by motional-averaging when the electron spin evolves quickly on the timescale of nuclear evolution.
At long times, the competition between coherent twisting dynamics, which squeezes fluctuations as $1/t$, and phase diffusion, which tends to increase fluctuations as $t^{1/2}$, slows down squeezing to $\widetilde{\Delta I} \sim t^{-1/2}$, but does not prevent it.

%%%%%%%%% DISCUSSION %%%%%%%%%
The results above are based on a semiclassical mean-field treatment of Eq.(\ref{eq:bloch_eqs}) supplemented by electron-spin-fluctuation-driven phase diffusion. % driven by electron spin fluctuations. %about the steady state given above 
This intuitive approach is quantitatively supported by a lengthier %more rigorous
 calculation based on the full density matrix of the combined electron-nuclear system, to be presented elsewhere\cite{Squeezing2}.
The more powerful density-matrix approach can also be used to study squeezing in the coherent (strong) driving regime of electron spin dynamics where the rate equations, Eq.(\ref{eq:rate_eqs}), cannot be applied.

\section{Discussion: Experimental Feasibility}

Throughout the discussion, we have worked within the approximation of uniform hyperfine coupling $A_n = A$, see Eq.(\ref{Hmicro}), which is widely used for studying electron-nuclear coupling in quantum dots.
More realistically, hyperfine coupling is strong near the center of the dot, where electron density is high, and weak at the edges.
% An analogous approximation of uniform coupling was also used to analyze squeezing in atomic ensembles where 
We note that similar variations in coupling occur in atomic ensembles when the size of the atom cloud is comparable to the wavelength of light\cite{Schleier-Smith,Schelier-SmithPRA}.
% Nonetheless, squeezing 
Squeezing has been demonstrated beautifully in that context, and thus it appears that the variation of couplings does not severely impact the effect.
% {\bf Lieven asks if Vladan doesn't have a better explanation than this.}

%For easily achieved polarizations of $5\%$, squeezing sets in after $t_s \sim 4 \mu$s, and fluctuations are suppressed by a factor of 10 within approximately $40 \mu$s (neglecting phase diffusion). 
For achievable polarizations of $20\%$, squeezing sets in after $t_S \sim 2\ \mu$s, and fluctuations are suppressed by a factor of 10 within approximately $20\ \mu$s (neglecting phase diffusion). 
% Note that due 
Due to classical fluctuations in the initial state, first $\frac1{\sqrt{p}}$-fold squeezing goes toward reaching the standard quantum limit, after which quantum squeezing proceeds. 
These timescales are $10-100$ times shorter than the approximately 1 ms-scale nuclear coherence time recently measured in vertical double quantum dots\cite{Takahashi10}. 
%\mpar{new ref. ok?} 
%$100$ $\mu$s-scale dipole-dipole dominated nuclear spin decoherence time that can be extracted from the NMR linewidth in bulk GaAs\cite{solidNMR} or from nuclear spin echo measurements in GaAs nanostructures\cite{Greilich2,Hirayama}. 
It should thus be possible to squeeze the nuclear spin state faster than it decoheres.

%Except for the twisting dynamics discussed in this paper,
All elements required for achieving and demonstrating squeezing have been realized experimentally in nanostructures at low temperature. 
Dynamical nuclear polarization can routinely be produced\cite{Baugh, Gammon} and controllably rotated using NMR pulses \cite{Machida1,Machida2,Hirayama,Tartakovskii, Takahashi10}, and
%We can pump the nuclei using dynamic nuclear polarization \cite{Baugh}, rotate the nuclear spin vector into the transverse plane with an NMR pulse \cite{Machida,Hirayama,Tartakovskii, Takahashi10}, rotate it back to the magnetic field axis after squeezing, and measure 
the degree of squeezing can be ascertained through electron spin dephasing measurements \cite{ Petta05, Mikkelsen, Greilich, Press10, Bluhm, Koppens08}.
%The driving requirements for achieving the twisting dynamics appear realistic as well. 
Coherent spin control has also been demonstrated in a variety of systems\cite{Koppens06, Mikkelsen, Press10}.
In paricular, we note that in Ref.~\cite{Koppens06} electron spin resonance was achieved by excitation using microwave magnetic fields, with driving amplitudes comparable to the random nuclear field acting on the electron spin, $A\delta I$.
%This corresponds to 
The corresponding 
transition rates $W$ are of the order of 10 MHz. 
In order to reach the motional averaging regime, the electron spin relaxation rate, $\Gamma_1$, must be comparable to the transition rate $W$, %at least five times higher than $W$. This is 
which can be easily accomplished by allowing cotunneling to the electron reservoirs next to the dot. 
%This rate is also faster than intrinsic decoherence processes for the electron spin, so $\gamma = 2 \Gamma_1$, as was assumed. [LV: WHY DID YOU WRITE $\gamma=4 \Gamma_1$ below eq (10)?]

%Quantum mechanical squeezing is achieved when $\langle (\delta \hat{I}^z)^2 \rangle < |\langle \hat{I}^x \rangle |/2$ \cite{Kitagawa}. For a spin bath with polarization $p$ along $x$ and variance $qN/4$ along $z$ after squeezing, the squeezing condition is $q<p$ ($q<3p/5$ for spin-3/2 nuclei). A $4\%$ polarization requires only a fivefold reduction in the uncertainty $\sqrt{\langle (\delta \hat{I}^z)^2 \rangle}$, which is easily possible experimentally.

\section{Conclusions}
In summary, squeezed states of nuclear spins can be produced in quantum dots via feedback provided by their hyperfine coupling to an electron spin driven close to resonance.
Our squeezing mechanism applies equally well to gate-defined dots subjected to microwave driving of the electron spin, and to optically controlled spins in self-assembled dot ensembles.
We have considered various physical effects that compete with squeezing, and estimated the timescale of squeezing.
Our estimates indicate that squeezing is feasible and can be realized with current capabilities.
%The squeezing mechanism \addMR{applies 
%, based on microwave driving of the electron spin, can also be easily adapted to the case of optically controlled single quantum dots or dot ensembles, where resonant Raman transitions are used to drive single electron spins. 
Such schemes open the door to unprecedented levels of quantum control over collective degrees of freedom in nanoscale systems with mesoscopic numbers $10^4$ to $10^6$ of nuclear spins.

\begin{acknowledgments}
We thank M. D. Lukin and the Delft spin qubit team for useful discussions. LV acknowledges the MIT Condensed Matter Theory group for its hospitality. This work was supported by the Dutch Foundation for Fundamental Research on Matter and a European Research Council Starting Grant (LV), the NSF-funded MIT-Harvard Center for Ultracold Atoms (LV and VV), the NSF grants DMR-090647 and PHY-0646094 (MR) and PHY-0855052 (VV), and IARPA (MR and LV).
 %, and the Delft spin qubit team for useful discussions.
\end{acknowledgments}

\appendix
\section{Squeezing of the Wigner Distribution}

In this section, we provide a mathematical description of squeezing by analyzing the evolution of a nuclear spin state characterized by a Gaussian Wigner distribution.
As discussed in the main text, for a large spin initially oriented in the $x$ direction, and for short times before the Wigner distribution extends significantly around the Bloch sphere, the Wigner distribution in a locally flat patch of Bloch sphere evolves as
%
%\be\label{eq:evolved_gaussian_app}
$f_t(I_y, I_z)=\mathcal{A} e^{-\frac12 {\bf v}^T Q {\bf v}}$, %\exp\lp - \frac{I_z^2+(I_y+I \lambda t I_z)^2}{2\delta I^2}\rp
%\ee
% 
see Eq.(10), %\ref{eq:evolved_gaussian}), 
with %$v = (I_y, I_z)^T$ and 
\be
\label{eq:quadform}
{\bf v} = \lp \begin{array}{c}I^y\\ I^z\end{array}\rp,\ \ \
Q = \frac{1}{\Delta I^2}\lp \begin{array}{cc}1 & \lambda I t\\\lambda I t & 1 + (\lambda I t)^2\end{array}\rp.
\ee
Here $\Delta I = \Delta I_0^{y,z}$ characterizes the transverse fluctuations in the initial nuclear spin state.

For times $t > 0$, the circular Wigner distribution is deformed to an ellipse, with major and minor axes determined by the quadratic form $Q$ in Eq.(\ref{eq:quadform}).
As shown in Fig.3 of the main text, stretching in one direction ($y'$) %$\cot \phi_t$ times 
is accompanied by squeezing in the perpendicular direction ($z'$), such that the
%by the same amount, preserving the
phase space volume of the Wigner distribution is preserved. 
The major and minor axes $y'$ and $z'$, which lie parallel to the eigenvectors of $Q$, are rotated relative to $y$ and $z$ by an angle $\theta$:
\be
\lp\begin{array}{c}I^{y'}\\I^{z'}\end{array}\rp = \lp\begin{array}{cc} \cos\theta &-\sin\theta\\ \sin\theta & \cos\theta\end{array}\rp \lp\begin{array}{c}I^y\\I^z\end{array}\rp.
\ee
The angle $\theta$ can be found by extremizing the quantity
\be
W = {\bf w}_\theta^T Q{\bf w}_\theta, \ \ {\bf w}_\theta = \lp\begin{array}{cc}\cos\theta\\ -\sin\theta\end{array}\rp
%I_z^2+(I_y+I \lambda t I_z)^2\propto \sin^2\theta+(\cos\theta +I \lambda t \sin\theta)^2
 %=1+I \lambda t \sin 2\theta+(I \lambda t)^2(1-\cos 2\theta)/2,
\ee 
%
%where $I_z/I_y=\sin\theta/\cos\theta$. 
%We obtain $\theta=-\phi_t$.
with respect to $\theta$.
Using the identity $[1 - \tan^2\theta]/2\tan\theta = \cot 2\theta$, we find 
\be\label{eq:theta}
\cot 2\theta = \lambda I t/2.
\ee
Note that Eq.(\ref{eq:theta}) has two solutions $\theta_{1,2}$ separated by 90$^{\circ}$, as expected for a symmetric form. % as expected for the $2\times 2$ symmetric matrix $Q$.
%, giving
%%
%The corresponding eigenvalues $q_{1,2}$ are given by
%\bea
%q_1 &\equiv& \frac{1}{\delta I_{y'}} = 1 + \frac12(\lambda I t)^2 - \frac12\sqrt{[2 + (\lambda I t)^2] - 4}\\
%q_2
%\eea

In the eigenbasis, we write
\be
\label{eq:wigner_twisting}
%f_t(I_z,I_y)=A \exp\lp -\frac{\cot^2\phi_t I_{z'}^2+\tan^2\phi_t I_{y'}^2}{2\delta I^2}\rp
f_t(I^{y'}, I^{z'}) = \mathcal{A} \exp\left[ -\frac12\lp\frac{I^{y'}}{\Delta I_+(t)}\rp^2 - \frac12\lp\frac{I^{z'}}{\Delta I_-(t)}\rp^2\right],
%\label{eq:stretching_factor}
%w_t =\sqrt{1+\frac{(\lambda I t)^2}{4}}+\frac{\lambda I t}{2}
%%w_t =\cot\phi_t 
%% ,\quad 
\ee
%
%where $\cot 2\phi_t=\lambda I t/2$. 
with 
\be
\Delta I^2_\pm(t) = \Delta I^2 \left(1 + \frac{ (\lambda I t)^2}{2}\left[1 \mp \sqrt{1 + \frac{4}{(\lambda I t)^2}}\right]\right)^{-1}.
\ee
In the long time limit $\lambda I t \gg 1$, the width $\widetilde{\Delta I}(t) \equiv \Delta I_-(t)$ of the squeezed component reduces to Eq.(11). %(\ref{dI_ideal}).

\section{Phase Diffusion}
The effect of time-dependent fluctuations of electron spin polarization about the mean field value can be analyzed within the rate equation model %, Eq.(\ref{eq:rate_eqs}), in the Supplementary Material.
by introducing a time-dependent quantity
\be
\tilde S^z(t)= S^z +\delta S^z(t)
.
\ee
The fluctuating part $\delta S^z$ can be modeled as delta-correlated noise
$\la \delta S^z(t')\delta S^z(t'')\ra\propto\delta(t'-t'')$, with an intensity determined by the rate process, Eq.(5). %(\ref{eq:rate_eqs}). 
%We obtain \mpar{\bf must check}
%
%\be
%\kappa =\frac{2W+\Gamma_1}{(W+\Gamma_1)W}
%.
%\ee
%
As shown in Supplementary Section \ref{sec:kappa}, %the effect of such noise can be described as phase diffusion,
such noise generates phase diffusion,
\be
\la \delta\theta^2(t)\ra = \kappa t, \quad \kappa = 2A^2\frac{(W+\Gamma_1)W}{(2W+\Gamma_1)^3},
\ee 
where $\delta\theta$ is the fluctuating part of the Larmor precession angle, $I^x+iI^y\propto e^{i(\theta + \delta \theta)}$. 
The phase diffusion can be accounted for by adding a diffusion term with diffusivity $\tilde\kappa= I^2\kappa $ to the equation describing the time evolution of the Wigner distribution. %, Eq.(\ref{eq:evolved_gaussian}).

An important consequence of phase diffusion is non-conservation of phase volume, which can be illustrated by the evolution of a Gaussian Wigner distribution.
Similar to the mean-field case, Eq.(10), %(\ref{eq:evolved_gaussian}), 
such a distribution evolves in time as
\be\label{eq:evolved_gaussian_diffusion}
f_t(I^y,I^z)=\mathcal{A}'(t) \exp\left[ -\frac{(I^z)^2}{2\Delta I^2} - \frac{(I^y+I \lambda t I^z)^2}{2(\Delta I^2+\tilde \kappa t)}\rp
,\quad t>0.
\ee
Initially, phase diffusion leads to a broadening of the Wigner distribution, characterized by the factor $\sqrt{1+\tilde\kappa t/\Delta I^2}$, which grows like $t^{1/2}$ for $\tilde\kappa t > \Delta I^2$.
At later times, $\Delta I\lambda t I \gtrsim \sqrt{\tilde \kappa t }$,  the behavior is dominated by the linear in $t$ twisting/stretching dynamics.
Therefore for times satisfying 
$t > t_{\rm noise} = \frac{2A^2\kappa}{N\lambda^2}$,
the coherent stretching overwhelms the effect of phase diffusion.

% The effect of noise is dominant at intermediate times, when phase diffusion broadens the distribution (\ref{eq:evolved_gaussian_diffusion}) more strongly than it is stretched by twisting.
% Therefore squeezing dominates over diffusion at long times, 
% %
% \be
% \delta I\lambda t I \gtrsim \sqrt{\tilde \kappa t }
% ,\quad
% \sqrt{\tilde \kappa t } \gtrsim \delta I 
% ,\quad
%  \delta I=\sqrt{N/2}
% \ee
% %
% This gives
% %
% \be
% \frac{N}{2 A^2 I^2 \kappa} \lesssim t\lesssim t_{\rm noise}=\frac{2A^2\kappa}{N\lambda^2}
% .
% \ee
% %
% For times longer than $t_{\rm noise}$, the effect of stretching, which is linear in $t$, overwhelms the effect of phase diffusion, growing only as $t^{1/2}$.

The efficiency of squeezing in the presence of phase diffusion can be estimated as follows. 
%Phase diffusion results in smearing of the Wigner distribution, characterized by a factor $\sqrt{1+\tilde\kappa t/\delta I^2}$. 
At long times $t\gg t_{\rm noise}$, the factor $\sqrt{1+\tilde\kappa t/\Delta I^2}$ describes an increase of the width of the Wigner distribution compared to its ideal squeezed value $\widetilde{\Delta I}$ in Eq.(11). %(\ref{dI_ideal}). %$\delta I\tan\phi_t \approx \delta I/(\lambda I t)$. 
Combining the $t^{1/2}$ smearing due to phase diffusion with the $t^{-1}$ squeezing,
we find that the width of the Wigner distribution decreases as $t^{-1/2}$  at long times:
\be
\widetilde{\Delta I}_{\rm noise} = \widetilde{\Delta I}(t)\sqrt{1+\tilde\kappa t/\Delta I^2}\approx \frac{\tilde\kappa^{1/2}}{\lambda I} t^{-1/2}
\ee
This expression describes the slowing of squeezing due to phase diffusion.

\section{Calculation of the Phase Diffusion Constant}\label{sec:kappa}

To analyze phase diffusion, we need to calculate the generating function for spin fluctuations driven by up-down and down-up switching. Denoting the two switching rates as $W$ and $W'$, we can obtain the generating function for spin fluctuations during the time interval $0<t'<t$ by approximating a continuous Poisson process by a discrete Markov process with a small time step $\Delta\ll  W^{-1},(W')^{-1},t$. We have
\bea
&&\chi(\lambda)=\lp\begin{array}{c}1\\1\end{array}\rp^{\rm T}\lb e^{i\Delta (\lambda/2)\sigma_3}R_\Delta
\rb^N \lp\begin{array}{c}1/2\\1/2\end{array}\rp
\\\nonumber
&& 
% A(\lambda)=\lp\begin{array}{cc}e^{i\Delta\lambda/2}&0\\0&e^{-i\Delta\lambda/2}\end{array}\rp
% ,\quad
R_\Delta=\lp\begin{array}{cc}1-W\Delta &W\Delta\\W'\Delta&1-W'\Delta\end{array}\rp
,\quad N=\frac{{t}}{\Delta}
,
\eea
where $W'=W+\Gamma_1$.
%$\Delta$ is a short time interval introduced for convenience (). 
Taking the limit $\Delta\to 0$, $N\to\infty$ we obtain an expression
\bea\label{eq:chi}
&&\chi(\lambda)=\lp\begin{array}{c}1\\1\end{array}\rp^{\rm T}
e^{M}
\lp \begin{array}{c}1/2\\1/2\end{array}\rp
,
\\
&& M={t}\lp \begin{array}{cc}i\lambda/2-W&W\\W'&-i\lambda/2-W'\end{array}\rp
.
\eea
The generating function (\ref{eq:chi}) provides a full description of the statistics of phase fluctuations, $\theta_{t}=\int_0^{t} S_Z(t)dt$, by encoding all its cumulants:
%. It can be used to obtain all cumulants of this quantity:
%
\be
\ln\chi(\lambda)=\sum_{k=1}^\infty m_k\frac{(i\lambda)^k}{k!}
,
\ee
with $m_1$ and $m_2$ giving the expectation value $\la S_z\ra {t}$ and the variance $\la(\theta_{t}-\la\theta_{t}\ra)^2\ra$, respectively. The latter quantity yields the phase diffusion constant via $m_2=\kappa {t}$.

Matrix exponential $e^M$ can be evaluated by writing it in terms of Pauli matrices, $M=x_0+x_i\sigma_i$, where
\be
x_0=-W_+
,\quad
x_1=W_+
,\quad
x_2=iW_-
,\quad
x_3=i\lambda/2-W_-
,
\ee
and we defined $W_\pm=(W\pm W')/2$. We have 
\be
e^M=e^{x_0{t}}\lp \cosh (X{t}) +\frac{\sinh (X{t})}{X}x_i\sigma_i\rp
\ee
where $X^2=x_1^2+x_2^2+x_3^2=W_+^2-\lambda^2/4-i\lambda W_-$. Plugging this expression for $e^M$ in Eq.(\ref{eq:chi}), we find
\be
\chi(\lambda)=2e^{x_0{t}}\lp \cosh (X{t}) +\frac{\sinh (X{t})}{X}x_1\rp
,
\ee
an exact expression which is valid both at short times and at long times.

To analyze fluctuations in the steady state, we focus on the long times $t\gg W^{-1},(W')^{-1}$. In this limit, the behavior of $\chi(\lambda)$ can be understood by replacing $\cosh Xt$ and $\sinh Xt$ by $e^{Xt}$, giving
% %
% \be
% \ln\chi(\lambda)= (X-W_+)t
% \ee
%
% To find the average value of spin polarization and its fluctuations, we analyze the quantity
% %
% \be
% \ln \chi(\lambda)= -W_+{t}+\ln \lp \frac12\lp 1+\frac{W_+}{X}\rp e^{X{t}}+\frac12\lp 1-\frac{W_+}{X}\rp e^{-X{t}}\rp
% %+\frac{e^{X{t}}-e^{-X{t}}}{X}W_+\rp
% \ee
% %
% % Since we are interested 
% in the long-limit limit of large $W^{-1}_+ t\gg 1$ (keeping $\lambda$ small). In this limit the expression for $\ln\chi$ is simplified:
%
\be
\ln \chi(\lambda)\approx (X-W_+){t}=-\frac{\lambda^2/4+i\lambda W_-}{X+W_+}{t}
\ee
Taylor expanding this expression up to order $\lambda^2$ we find the first and second cumulants of phase fluctuations:
% , giving the time-averaged polarization and the phase diffusion constant $\kappa$:
%
\be
\ln\chi(\lambda)=-i\lambda\frac{W_-{t}}{2W_+}+\frac{(i\lambda)^2}2\frac{(W_+^2-W_-^2){t}}{4W_+^3}+O(\lambda^3)
\ee
Substituting $W'=W+\Gamma_1$, we obtain the time-averaged polarization and the phase diffusion constant
\be
\la S_z\ra=\frac12\frac{\Gamma_1}{2W+\Gamma_1}
,\quad
\kappa =2\frac{(W+\Gamma_1)W}{(2W+\Gamma_1)^3}
\ee
% $\ln\chi$: 
%
Crucially, the phase diffusion slows down when the switching rates $W$ and $W'$ grow, which justifies our motional averaging approximation.

\end{document}